\documentclass[conference]{IEEEtran}
\IEEEoverridecommandlockouts
\usepackage{cite}
\usepackage{amsmath,amssymb,amsfonts}
\usepackage{algorithmic}
\usepackage{graphicx}
\usepackage{textcomp}
\usepackage{xcolor}
\def\BibTeX{{\rm B\kern-.05em{\sc i\kern-.025em b}\kern-.08em
    T\kern-.1667em\lower.7ex\hbox{E}\kern-.125emX}}
\begin{document}

\title{The road to Sustainable DevOps\\
}
\author{\IEEEauthorblockN{Darwish Ahmad Herati, Maria Clara Aderne, Fabio Kon}
\IEEEauthorblockA{\textit{Department of Computer Science} \\
\textit{IME - University of São Paulo}\\
\{darwish,kon\}@ime.usp.br}
}

\maketitle

\begin{abstract}

This manuscript focuses on the environmental, social, and individual sustainability dimensions within the modern software development lifecycle, aiming to establish a holistic approach termed Sustainable DevOps (SusDevOps). Moving beyond the already well-researched economic and technical aspects, our approach to SusDevOps emphasizes the importance of minimizing \emph{environmental} impacts, fostering social inclusion, and supporting individual well-being in software engineering practices. We highlight some key challenges in incorporating these dimensions, such as reducing ecological footprints, promoting workforce inclusion, and addressing the individual well-being of developers. We plan to adopt a structured approach incorporating systematic literature reviews, surveys, and interviews to deepen our understanding, identify gaps, and evolve actionable, sustainable practices within the DevOps community. Collectively, these initiatives can contribute to a more sustainable software engineering ecosystem.

\end{abstract}


\section{Introduction}
Software engineering can be seen as the organized application of scientific and technological expertise, techniques, and experience in software design, implementation, testing, and documentation, to improve production, support, and quality \cite{ISO2382}.  It is a systematic, disciplined and measurable approach to the development, operation, and maintenance of software, effectively incorporating engineering principles into software practices \cite{IEEE1990SET}.


Within this field, DevOps has emerged as a transformative approach that merges development (Dev) and operations (Ops) practices. It brings together multiple disciplines to streamline and automate the continuous delivery of software updates. This effort emphasizes collaboration across teams to maintain high standards of accuracy and reliability in each release \cite{leite2019survey}.


In contrast, the United Nations defined \emph{sustainable development} as the process of fulfilling the needs of the current generation while ensuring that future generations can also meet their own needs \cite{UNWCED1987}.
Sustainability is often illustrated through the three pillars of social, economic, and environmental dimensions, typically depicted as intersecting circles with sustainability at their core \cite{purvis2019three}. An examination of historical literature reveals that this three-pillar framework did not emerge from a single source but developed gradually from critiques of the economic status quo, influenced by social and ecological perspectives. Additionally, the effort by the UN to harmonize economic growth with social and ecological challenges contributed to its formation. Despite the widespread use of the three-pillar model, a rigorous theoretical foundation remains elusive due to the diverse schools of thought within the sustainability discourse. This lack of a solid theoretical framework poses challenges for effectively operationalizing the concept of sustainability.

\section{Sustainability in Software Engineering}

In the context of software engineering, researchers have expanded the idea of sustainability to address the unique demands of technology, introducing five distinct dimensions~\cite{becker2015sustainability}:

\begin{itemize}
    \item \textbf{Environmental:} addresses the long-term impacts of human activities on natural systems, covering aspects such as ecosystems, resource use, climate change, food production, water, pollution, and waste management.
    \item \textbf{Social:} focuses on societal communities and the factors that diminish trust, encompassing equity, justice, employment, and democracy.
    \item \textbf{Economic:} centered on assets and capital, including wealth generation, profitability, investment, and income. 
    \item \textbf{Technical:} dealing with the durability of information, systems, and infrastructure, emphasizing maintenance, innovation, obsolescence, and data integrity in the context of changing conditions.
    \item \textbf{Individual:} related to the well-being of individuals, considering mental and physical health, education, self-esteem, skills development, and mobility.
\end{itemize}


Economic sustainability has emerged as a paramount concern for companies, reflecting the increasing recognition of its critical role in long-term viability and competitiveness. For almost a century, academic research has delved into various facets of economic sustainability, exploring methodologies that enable organizations to minimize operational costs while maximizing profitability and efficiently managing resources.


Technical sustainability has emerged a significant concern within the field of software engineering and has been the focus of academic research for the past three decades. This area of study addresses the need of software systems to remain functional and relevant over time as they evolve and adapt to changing requirements, advancements in technology, and shifting user needs. Researchers have investigated various strategies to ensure that software can be efficiently maintained, including methodologies for refactoring code, improving software architecture, and adopting agile practices that facilitate iterative development and continuous feedback. Also, the importance of documentation, version control, and robust testing frameworks has been emphasized to enhance the long-term viability of software systems. As digital transformation accelerates across industries, the ongoing pursuit of technical sustainability remains crucial, ensuring that software not only meets current demands, but is also resilient enough to adapt to future challenges and opportunities.


While economic and technical sustainability are well-established, areas in software engineering, the dimensions of social, environmental, and individual sustainability remain relatively unexplored. Social sustainability focuses on ensuring equity, inclusion within software design, development processes, and the workforce, promoting accessibility for all users and fostering diverse participation in technology creation. 

This encompasses ethical considerations, human values, and the broader societal implications of software systems, which can significantly influence user experiences and societal norms~\cite{moises2023social}.

Environmental sustainability addresses the ecological impact of software systems, from development processes to deployment and software usage, while considering the energy consumption, environmental impact measurements, and resource utilization.


Souza et al. \cite{moises2023social} reviewed 19 studies, highlighting the importance of considering social factors, such as accessibility, inclusion, diversity, equity, and ethical values in creating sustainable software that positively impacts society. Although sustainable software development emphasizes environmental, economic, and social dimensions, there is a shortage of clear methods to address social factors specifically.


Penzenstadler \cite{penzenstadler2020attention}  defines individual sustainability in software engineering as the upkeep of personal human capital, which includes aspects such as health, education, skills, knowledge, leadership, and access to services. Moreover, it emphasizes the necessity of a strong, connected movement of citizens to create a sustainable, equitable, and democratic society. It advocates for empowering individuals, particularly software engineers, to maintain their physical, mental, and emotional health through proven methods that align with precautionary principles by addressing the risk of unsustainable practices among software developers. The goal is to reduce stress and the prevalence of depression and anxiety in the profession. It also highlights the importance of self-care practices, such as breathing and movement exercises, to help software professionals care for themselves and others.

\section{Towards SusDevOps}

Our research group focuses on the promotion of Sustainable DevOps (SusDevOps) as a forward-looking approach in software engineering. Inspired by previous work both within and outside our group \cite{becker2015sustainability, leite2019survey} we define: \textbf{SusDevOps} integrates sustainability into the DevOps process, automating reliable software delivery while considering environmental impact, individual well-being, social responsibility, economic viability, and technical efficiency. SusDevOps aims to balance continuous delivery with long-term sustainability across all these dimensions.


Despite the efforts, the community faces significant research challenges: the absence of a common understanding of the concept of sustainability and how it relates to software development;  the lack of metrics to evaluate the sustainability of software in distinct domains; and the need to evaluate the methods that aim to resolve the sustainability problems as well as their impacts over the years \cite{oyedeji2024integratingsustainabilityconcernsagile,venters2023sustainable}.

Atadoga et al. \cite{atadoga2024tools} note that the software industry is increasingly concerned with minimizing its environmental footprint, focusing on energy use, resource efficiency and carbon emissions. However, we believe that most of this trend is due to its positive economic impact and, in a much smaller extent, to ESG practices that benefit the company's image in the market. 

A much stronger approach towards environmental sustainability is needed if we want to avoid the growing negative impact of the IT industry on the planet. The way to achieve this is not clear. Improved regulation, mandatory laws, and novel sustainability-aware development methodologies and tools are potential solutions.

Academics can employ a range of research methods to address these challenges. In particular, our research group has started to gather information on the topic to build a basic body of knowledge that can help us tackle these challenges.

Systematic literature reviews of existing research related to environmental sustainability in software engineering can help us understand what is the current body of knowledge in this area, helping to synthesize insights, identify gaps, and establish a theoretical framework for understanding the intersections of these challenges. 

Conducting surveys and questionnaires with industry experts can be effective in gathering data on software engineers' awareness, attitudes, and behaviors regarding sustainability practices. This can provide qualitative insights into common barriers and motivations among practitioners and help to assess the level of awareness of sustainable practices. Interviews with software engineers will help to capture their personal experience and expectations with regard to sustainability.

Then, as a consequence, the scientific community should focus on developing theories, conceptual frameworks, methods, guidelines, and techniques to improve the sustainability of the development and operation of our software systems. Finally, both academics and the industry should work towards the development of automated tools to incorporate sustainability principles in all stages of the DevOps lifecycle.


\section{Conclusion}

There is an urgent need for incorporating sustainability into software engineering, by shifting the focus from the economic and technical dimensions to include environmental, social, and individual dimensions. As software development continues to evolve, the integration of sustainability into DevOps through SusDevOps can bring a holistic approach that balances efficient software delivery with sustainable practices across all dimensions. The key challenges in this integration involve mitigating environmental impacts, promoting social inclusion, and improving individual well-being. Collectively, these efforts aim to transform the industry towards greater sustainability.

\bibliographystyle{IEEEtran}
\bibliography{References}

\begin{thebibliography}{10}
\providecommand{\url}[1]{#1}
\csname url@samestyle\endcsname
\providecommand{\newblock}{\relax}
\providecommand{\bibinfo}[2]{#2}
\providecommand{\BIBentrySTDinterwordspacing}{\spaceskip=0pt\relax}
\providecommand{\BIBentryALTinterwordstretchfactor}{4}
\providecommand{\BIBentryALTinterwordspacing}{\spaceskip=\fontdimen2\font plus
\BIBentryALTinterwordstretchfactor\fontdimen3\font minus
  \fontdimen4\font\relax}
\providecommand{\BIBforeignlanguage}[2]{{%
\expandafter\ifx\csname l@#1\endcsname\relax
\typeout{** WARNING: IEEEtran.bst: No hyphenation pattern has been}%
\typeout{** loaded for the language `#1'. Using the pattern for}%
\typeout{** the default language instead.}%
\else
\language=\csname l@#1\endcsname
\fi
#2}}
\providecommand{\BIBdecl}{\relax}
\BIBdecl

\bibitem{ISO2382}
{International Organization for Standardization}, ``Information technology —
  vocabulary — part 1: Fundamental terms,'' 1993, iSO/IEC 2382-1:1993.

\bibitem{IEEE1990SET}
``Ieee standard glossary of software engineering terminology,'' \emph{IEEE Std
  610.12-1990}, pp. 1--84, 1990.

\bibitem{leite2019survey}
L.~Leite, C.~Rocha, F.~Kon, D.~Milojicic, and P.~Meirelles, ``A survey of
  devops concepts and challenges,'' \emph{ACM Computing Surveys (CSUR)},
  vol.~52, no.~6, pp. 1--35, 2019.

\bibitem{UNWCED1987}
{United Nations Secretary-General} and {World Commission on Environment and
  Development}, ``Report of the world commission on environment and
  development: Note,'' New York, 1987, symbol A/42/427.

\bibitem{purvis2019three}
B.~Purvis, Y.~Mao, and D.~Robinson, ``Three pillars of sustainability: in
  search of conceptual origins,'' \emph{Sustainability science}, vol.~14, pp.
  681--695, 2019.

\bibitem{becker2015sustainability}
C.~Becker, R.~Chitchyan, L.~Duboc, S.~Easterbrook, B.~Penzenstadler, N.~Seyff,
  and C.~C. Venters, ``Sustainability design and software: The karlskrona
  manifesto. in 2015 ieee/acm 37th ieee international conference on software
  engineering (vol. 2, pp. 467-476),'' 2015.

\bibitem{moises2023social}
A.~C. Moises~de Souza, D.~Soares~Cruzes, L.~Jaccheri, and J.~Krogstie, ``Social
  sustainability approaches for software development: a systematic literature
  review,'' in \emph{International Conference on Product-Focused Software
  Process Improvement}.\hskip 1em plus 0.5em minus 0.4em\relax Springer, 2023,
  pp. 478--494.

\bibitem{penzenstadler2020attention}
B.~Penzenstadler, ``Where attention goes, energy flows: enhancing individual
  sustainability in software engineering,'' in \emph{Proceedings of the 7th
  International Conference on ICT for Sustainability}, 2020, pp. 139--146.

\bibitem{oyedeji2024integratingsustainabilityconcernsagile}
\BIBentryALTinterwordspacing
S.~Oyedeji, R.~Chitchyan, M.~O. Adisa, and H.~Shamshiri, ``Integrating
  sustainability concerns into agile software development process,'' 2024.
  [Online]. Available: \url{https://arxiv.org/abs/2407.17426}
\BIBentrySTDinterwordspacing

\bibitem{venters2023sustainable}
C.~C. Venters, R.~Capilla, E.~Y. Nakagawa, S.~Betz, B.~Penzenstadler, T.~Crick,
  and I.~Brooks, ``Sustainable software engineering: Reflections on advances in
  research and practice,'' \emph{Information and Software Technology}, p.
  107316, 2023.

\bibitem{atadoga2024tools}
A.~Atadoga, U.~J. Umoga, O.~A. Lottu, and E.~O. Sodiy, ``Tools, techniques, and
  trends in sustainable software engineering: A critical review of current
  practices and future directions,'' \emph{World Journal of Advanced
  Engineering Technology and Sciences}, vol.~11, no.~1, pp. 231--239, 2024.

\end{thebibliography}
\end{document}